\documentclass[12pt]{iopart}
\usepackage{graphicx}
\begin{document}
\title[Transfer of entanglement from electrons to photons]{\Large Transfer of entanglement from electrons to photons by optical selection rules}
\author{M. Titov\dag, B. Trauzettel\ddag, B. Michaelis\ddag,\\
and C.W.J. Beenakker\ddag}
\address{\dag\ Max-Planck-Institut f\"{u}r Physik komplexer Systeme,\\
N\"{o}thnitzer Str.\ 38, 01187 Dresden, Germany}
\address{\ddag\ Instituut-Lorentz, Universiteit Leiden,\\
P.O. Box 9506, 2300 RA Leiden, The Netherlands}

\begin{abstract}
The entanglement transfer from electrons localized in a pair of quantum dots to circularly polarized photons is governed by optical selection rules, enforced by conservation of angular momentum. We point out that the transfer can not be achieved by means of unitary evolution unless the angular momentum of the two initial qubit states differs by $2$ units of $\hbar$. In particular, for spin-entangled electrons the difference in angular momentum is $1$ unit --- so the transfer fails. Nevertheless, the transfer can be successfully completed if the unitary evolution is followed by a measurement of the angular momentum of each quantum dot and post-processing of the photons using the measured values as input.
\end{abstract}
\submitto{New Journal of Physics, focus issue on ``Solid State Quantum Information''}

\section{Introduction}

A key step in road maps for solid-state quantum information processing is the transfer of an entangled state from localized to flying qubits and vice versa. Several different schemes exist for the transfer of entanglement from squeezed radiation to localized qubits, e.g.\ distant atoms or superconducting quantum interference devices \cite{Kra04,Pat04a,Pat04b,Pat04c}. In another class of proposals, the entanglement is transferred from the localized qubits of electron spins to the flying qubits of photon polarizations \cite{Cer04}. {\em Classical\/} correlations between the spins can be transferred to the polarizations when conservation of angular momentum together with spin-orbit coupling imposes a one-to-one relation between the spin of the electron and the polarization of the photon that it produces in a radiative transition. Entanglement, however, is a quantum correlation which is easily lost in this process.

The obstacle to entanglement transfer is that the optical selection rules in the general case entangle the photons with the electrons --- and then the entanglement of the photons among themselves is lost once one traces out the electronic degrees of freedom \cite{Vrij01}. This ``tracing out'' is unavoidable when the photon state is measured independently of the electron state. After explaining this difficulty in some more detail, we will show that it can be circumvented by post-processing the photon state with the input of information obtained from a measurement on the electron state.

Our analysis has certain implications for a recent realistic proposal by Cerletti, Gywat, and Loss \cite{Cer04} to use electron-hole recombination in spin light-emitting diodes (spin-LEDs \cite{Fie99,Ohn99}) as an efficient method for the transfer of entanglement from electron spins onto circular photon polarizations. We will argue, firstly, that the method of Ref.\ \cite{Cer04} transfers classical correlations but not quantum correlations; and, secondly, that the quantum entanglement transfer can be realized by measurement in a rotated basis of the hole angular momentum in each quantum dot after the photo-emission, followed by a single-photon operation conditioned on the outcome of that measurement.

In the concluding section we briefly discuss alternative schemes for quantum entanglement transfer, which do not require the
subsequent measurements (post-processing) and might, therefore, be realized more easily in the laboratory.

\section{General analysis}

We consider two quantum dots $A$ and $B$, each containing one qubit. The initial two-qubit electronic state has the generic form
\begin{equation}
|\Psi_{\rm in}\rangle= \alpha |\downarrow\downarrow \rangle+\beta |\uparrow\uparrow \rangle,\;\; 
|\alpha|^2+|\beta|^2=1.\label{Psiin}
\end{equation}
The entanglement of formation of this state is quantified by the concurrence \cite{Woo98} 
\begin{equation}
{\cal C}_{\rm in}=2|\alpha\beta|.
\end{equation}
The two states $\downarrow$, $\uparrow$ of the qubits are eigenstates of the total (orbital $+$ spin) angular momentum operator ${\cal L}_{z}$ in the $z$-direction, with eigenvalues $L_{\downarrow}$, $L_{\uparrow}$ (in units of $\hbar$). The first state in the ket $|\cdot\cdot\rangle$ refers to the qubit in quantum dot $A$ and the second state refers to quantum dot $B$.

Photons with opposite circular polarizations $\sigma_{\pm}$ (angular momentum $\pm 1$), emitted along the spin quantization axis, are produced according to the unitary evolution
\numparts
\begin{eqnarray}
&&|\downarrow\rangle|0\rangle\mapsto |\Phi_{+}\rangle|\sigma_+ \rangle,\\
&&|\uparrow\rangle|0\rangle\mapsto |\Phi_{-}\rangle|\sigma_- \rangle,
\end{eqnarray}
\endnumparts
where $|0\rangle$ denotes the photon vacuum and $|\Phi_{\pm}\rangle$ denotes the state of the quantum dot after the photo-emission of a $\sigma_{\pm}$ photon. Conservation of angular momentum requires that $|\Phi_{+}\rangle$ and $|\Phi_{-}\rangle$ are eigenstates of ${\cal L}_{z}$ with eigenvalues
\begin{equation}
L_{+}=L_{\downarrow}-1,\;\;L_{-}=L_{\uparrow}+1,
\end{equation}
respectively. In general, the two states $|\Phi_{+}\rangle$ and $|\Phi_{-}\rangle$ are orthogonal because they correspond to different eigenvalues $L_{+}\neq L_{-}$. The exception is the special case $L_{\downarrow}-L_{\uparrow}= 2$, when $L_{+}=L_{-}$ so that $|\Phi_{+}\rangle$ and $|\Phi_{-}\rangle$ may have a nonzero overlap.

The final state 
\begin{equation}
|\Psi_{\rm final}\rangle=\alpha|\Phi_{+}\Phi_{+}\rangle|\sigma_+\sigma_+\rangle+
\beta|\Phi_{-}\Phi_{-}\rangle|\sigma_-\sigma_-\rangle \label{Psifinal}
\end{equation}
represents an {\em encoding\/} rather than a {\em transfer\/} of the entanglement. Assuming that the photons are measured independently of the electrons, we trace out the electronic degrees of freedom to obtain the reduced density matrix $\rho_{\rm photon}$ of the photons by themselves:
\begin{eqnarray}
\fl\rho_{\rm photon}&=&{\rm Tr}_{\rm electron}\,|\Psi_{\rm final}\rangle\langle\Psi_{\rm final}|\nonumber\\
\fl&=&|\alpha|^2|\,\sigma_+\sigma_+\rangle\langle \sigma_+\sigma_+|+
|\beta|^2|\,\sigma_-\sigma_-\rangle\langle \sigma_-\sigma_-|+\gamma|\sigma_+\sigma_+\rangle\langle\sigma_-\sigma_-|+
\gamma^{\ast}|\sigma_-\sigma_-\rangle\langle\sigma_+\sigma_+|,\nonumber\\
\fl&&\label{rhophoton}
\end{eqnarray}
where $\gamma=\alpha \beta^{\ast}
\langle\Phi_{-}|\Phi_{+}\rangle_A\langle\Phi_{-}|\Phi_{+}\rangle_B$.
The concurrence of the mixed state $\rho_{\rm photon}$ is 
\begin{equation}
{\cal C}_{\rm final}=2|\gamma|.
\end{equation}
If $L_{\downarrow}-L_{\uparrow}\neq 2$, so that the final electronic states $|\Phi_+\rangle_X$ and $|\Phi_-\rangle_X$ in quantum dots $X=A, B$ are orthogonal, the polarizations of the photons have been correlated but not entangled ($\gamma=0\Rightarrow{\cal C}_{\rm final}=0$). Since unitary operations on the electronic degrees of freedom do not change $\rho_{\rm photon}$, the entanglement can not be recovered by unitary evolution once the photons have left the quantum dots and their evolution has decoupled from the electrons.

While unitary evolution can not disentangle the electrons from the photons, a projective measurement of the quantum dots followed by post-processing of the photons can realize the entanglement transfer. Considering the generic case $L_{\downarrow}-L_{\uparrow}\neq 2$, so that $\langle\Phi_{-}|\Phi_{+}\rangle=0$, we first perform the following local unitary operation on each of the two quantum dots:
\numparts
\begin{eqnarray}
&&|\Phi_{+}\rangle\mapsto\bigl(|\Phi_{+}\rangle+|\Phi_{-}\rangle\bigr)/\sqrt{2},\\
&&|\Phi_{-}\rangle\mapsto\bigl(|\Phi_{+}\rangle-|\Phi_{-}\rangle\bigr)/\sqrt{2}.
\end{eqnarray}
\endnumparts
We then measure ${\cal  L}_{z}$. The outcome of the measurement on dot $X=A,B$ is denoted by $L_X$. The measurement leaves the photons in the state
\begin{equation}
|\Psi_{\rm photon}\rangle=\alpha|\sigma_+\sigma_+\rangle+(-1)^{x}\beta|\sigma_{-}\sigma_{-}\rangle,\;\;
x=\frac{L_{A}-L_{B}}{L_{+}-L_{-}}. \label{Psimeasured}
\end{equation}
If the measurement gives $L_{A}=L_{B}$ no post-processing is needed; otherwise, the conditional phase shift $|\sigma_{\pm}\rangle\mapsto\pm|\sigma_{\pm}\rangle$ performed on one of the two photons completes the entanglement transfer.

\section{Application to spin-LEDs}

\begin{figure}
\includegraphics[width=8cm]{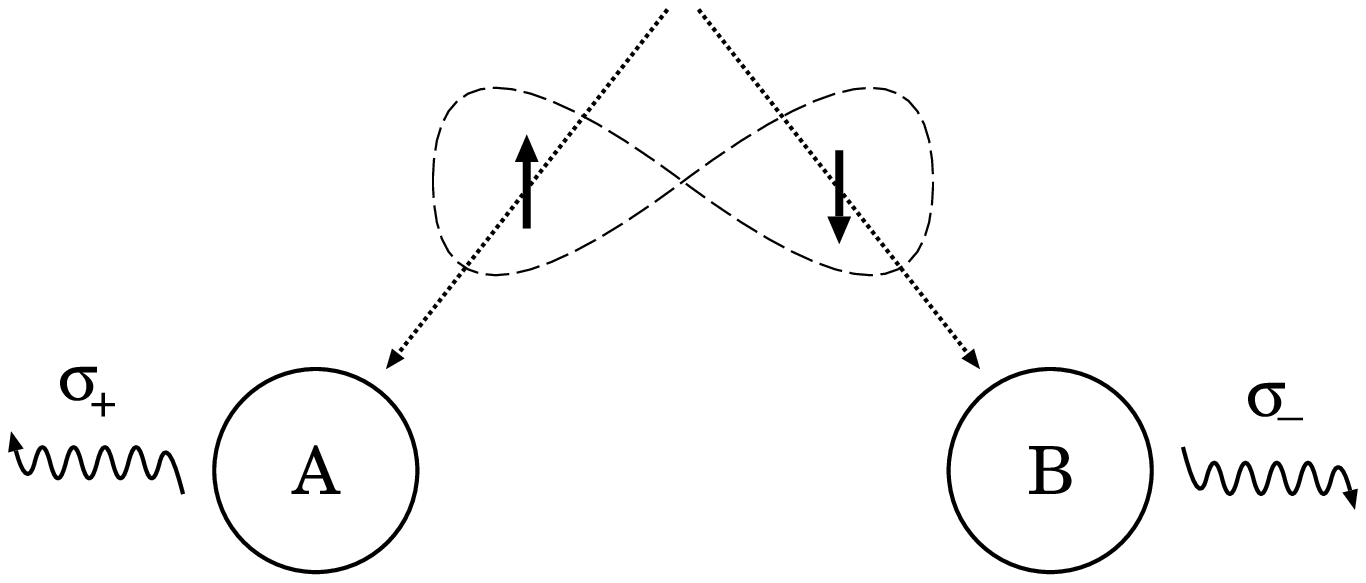}\qquad
\includegraphics[width=6cm]{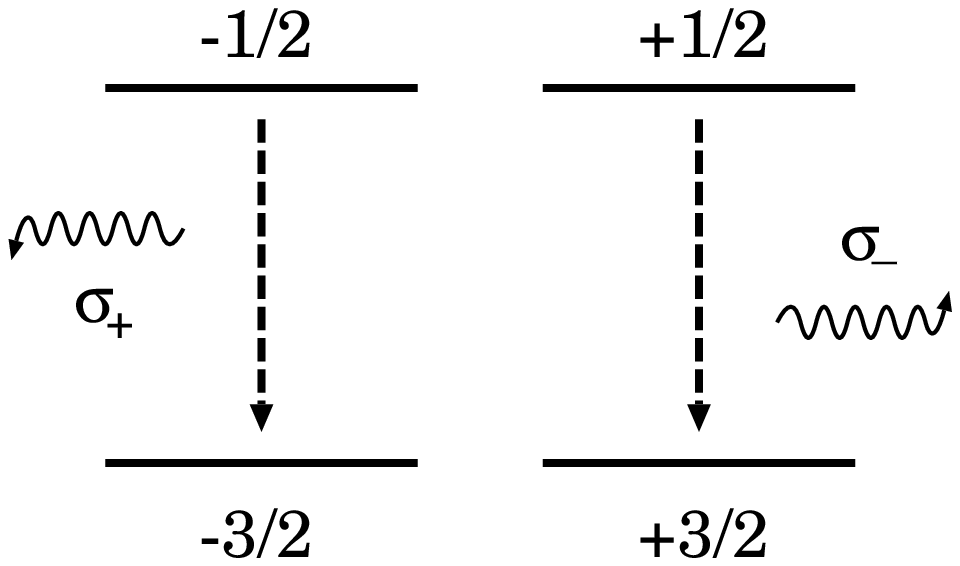}
\caption{Left panel: A spin-entangled pair of electrons recombines with a hole in quantum dots $A$ and $B$, to emit a pair of photons with anti-correlated circular polarizations $\sigma_{\pm}$. Right panel: The four lowest energy levels involved in the photo-emission of each quantum dot. The angular momentum quantum number is indicated. Initially both lower levels are filled by heavy holes. The recombination with a single electron in one of the two upper levels leaves the remaining hole entangled with the emitted photon. This prevents the transfer of the entanglement from the electrons to the photons.
}
\label{spinLED}
\end{figure}

The mechanism for entanglement transfer in spin-LEDs proposed in Ref.\ \cite{Cer04} is shown schematically in Fig.\ \ref{spinLED}. Two spin-entangled electrons (spin $\pm 1/2$) are injected into the conduction band of two different quantum dots, each of which is charged with a pair of heavy holes in the valence band. The two heavy holes in each quantum dot have opposite angular momentum $\pm 3/2$, so that their total angular momentum along $z$ vanishes. The initial state $|\Psi_{\rm in}\rangle$ is of the form (\ref{Psiin}), with the identification $|\uparrow\rangle\equiv|+\frac{1}{2},+\frac{3}{2},-\frac{3}{2}\rangle$ and $|\downarrow\rangle\equiv|-\frac{1}{2},+\frac{3}{2},-\frac{3}{2}\rangle$. (The three fractions indicate the angular momentum quantum numbers of the electron and the two heavy holes.) Electron-hole recombination can proceed either from angular momentum $+1/2$ to $+3/2$ with emission of a $\sigma_{-}$ photon or from $-1/2$ to $-3/2$ with emission of a $\sigma_{+}$ photon.

The remaining heavy holes become entangled with the photons, so that the final state $|\Psi_{\rm final}\rangle$ is of the form (\ref{Psifinal}) with the identification $|\Phi_{+}\rangle\equiv|+\frac{3}{2}\rangle$ and $|\Phi_{-}\rangle\equiv|-\frac{3}{2}\rangle$. These two states refer to two heavy holes with opposite angular momentum, so they are definitely orthogonal. Hence the reduced density matrix of the photons $\rho_{\rm photon}$ is of the form (\ref{rhophoton}) with $\gamma=0$ and the concurrence ${\cal C}_{\rm final}=0$. The polarizations of the photons have become correlated but not entangled. No matter how the remaining holes evolve after the photons have decoupled, the degree of entanglement of $\rho_{\rm photon}$ remains zero.

As explained in the previous section, the holes can be disentangled from the photons by post-processing in a sequence of three steps:
\begin{enumerate}
\item Bring each heavy hole in a superposition of states with opposite angular momentum by means of the local unitary operation $|\pm\frac{3}{2}\rangle\mapsto \bigl(|+\frac{3}{2}\rangle\pm|-\frac{3}{2}\rangle\bigr)/\sqrt{2}$.
\item Measure the angular momentum of each hole in the $z$-direction, with outcome $L_{A},L_{B}$.
\item Perform the conditional phase shift $|\sigma_{\pm}\rangle\mapsto\pm|\sigma_{\pm}\rangle$ on one of the two photons if $L_{A}\neq L_{B}$. 
\end{enumerate}
Step three is a routine linear optical operation. Step two might be achieved by detecting whether or not a spin-up heavy hole (angular momentum $+3/2$) can be injected separately into each of the two quantum dots. If both heavy holes enter their quantum dot, or if both do not enter, then $L_{A}=L_{B}$, while $L_{A}\neq L_{B}$ if one hole enters and the other does not. Step one might be achieved by an optical Raman transition \cite{Ima99,Cer05}.

\section{Conclusion}

We have shown that the transfer of entanglement from localized electron spins to circular photon polarizations by means of optical selection rules can not be achieved solely by unitary evolution. Projective measurements and post-processing conditioned on the outcome of the measurements are required as well, to disentangle the final electronic state from the photons. This difficulty originates from the mismatch between the half-integer spin of fermions and the integer spin of bosons. It severly complicates the original spin-LED proposal of Cerletti, Gywat, and Loss \cite{Cer04}, see Ref.\ \cite{Cer05}. In this concluding section we discuss several strategies that one might use to avoid the difficulty.

As proposed by Vrijen and Yablonovitch \cite{Vrij01}, entanglement transfer by unitary evolution to {\em linearly\/} polarized photons is possible if a strong magnetic field lifts the degeneracy between the up and down hole spins, so that the topmost hole state is nondegenerate. In the case of circular polarization considered here, it is possible if the angular momentum difference of the initial electronic qubits satisfies $L_{\downarrow}-L_{\uparrow}=2$. This might apply to a qubit formed from a $+3/2$ heavy hole and a $-1/2$ light hole. (The difference in mass might well prevent the formation of an entangled pair out of these qubits.) The spin-LEDs would then initially each contain a single $+1/2$ electron, which would recombine with the hole under emission of a $\sigma_{\pm}$ photon. The unique final state in this case is a pair of empty quantum dots. 

Alternatively, one might construct a qubit solely out of orbital degrees of freedom (without spin-orbit coupling): An electron in a circularly symmetric quantum dot has degenerate eigenstates of orbital angular momentum $+1$ or $-1$, which would decay to the nondegenerate ground state (zero angular momentum) with emission of a $\sigma_{\pm}$ photon \cite{Ema05}. Since the final state of the quantum dot is unique, it is not entangled with the photons, in accord with the general condition $L_{\downarrow}-L_{\uparrow}=2$ for the transfer of entanglement by unitary evolution.

An alltogether different way out of the constraints imposed by the optical selection rules is to let the spin-entangled electrons recombine with an {\em entangled\/} pair of holes. More specifically, if a pair of electrons in the state $\alpha|+\frac{1}{2},-\frac{1}{2}\rangle+\beta|-\frac{1}{2},+\frac{1}{2}\rangle)$ recombines with a pair of heavy holes in the singlet state $(|+\frac{3}{2},-\frac{3}{2}\rangle-|-\frac{3}{2},+\frac{3}{2}\rangle)/\sqrt{2}$, then the final photonic state (after tracing out the electronic degrees of freedom) becomes
\begin{equation}
\rho_{\rm photon}=\frac{1}{2}|\Psi\rangle\langle\Psi|+\frac{1}{2}|
0\rangle\langle 0|,\;\;
|\Psi\rangle=\alpha|\sigma_{-}\sigma_{+}\rangle-\beta|
\sigma_{+}\sigma_{-}\rangle,
\end{equation}
where $|0\rangle$ denotes the photon vacuum state. Detection of the photon pair projects onto the entangled state $|\Psi\rangle$. The efficiency of this entanglement transfer scheme is $1/2$ rather than unity, but it has the advantage that no measurement on the electronic state needs to be performed.

\ack
We have benefitted from discussions with M. Blaauboer and L. P. Kouwenhoven. This work was supported by the Dutch Science Foundation NWO/FOM.


\section*{References}

\begin{thebibliography}{99}
\bibitem{Kra04} B. Kraus and J. I. Cirac, Phys.\ Rev.\ Lett.\ {\bf 92}, 013602 (2004).
\bibitem{Pat04a} M. Paternostro, W. Son, and M. S. Kim, Phys.\ Rev.\ Lett.\ {\bf 92}, 197901 (2004).
\bibitem{Pat04b} M. Paternostro, W. Son, M. S. Kim, G. Falci, and G. M. Palma, Phys.\ Rev.\ A {\bf 70}, 022320 (2004).
\bibitem{Pat04c} M. Paternostro, G. Falci, M. S. Kim, and G. M. Palma, Phys.\ Rev.\ B {\bf 69}, 214502 (2004).
\bibitem{Cer04} V. Cerletti, O. Gywat, and D. Loss, cond-mat/0411235, version 1.
\bibitem{Vrij01} R. Vrijen and E. Yablonovitch, Physica E {\bf 10}, 569 (2001).
\bibitem{Fie99} R. Fiederling, M. Keim, G. Reuscher, W. Ossau, G. Schmidt, A. Waag, and L. W. Molenkamp, Nature {\bf 402}, 787 (1999).
\bibitem{Ohn99} Y. Ohno, D. K. Young, B. Beschoten, F. Matsukura, H. Ohno, and D. D.  Awschalom, Nature {\bf 402}, 790 (1999).
\bibitem{Woo98} W. K. Wootters, Phys.\ Rev.\ Lett.\ {\bf 80}, 2245 (1998).
\bibitem{Ima99} A. Imamoglu, D. D. Awschalom, G. Burkard, D. P. DiVincenzo, D. Loss, M. Sherwin, and A. Small, Phys.\ Rev.\ Lett.\ {\bf 83}, 4204 (1999).
\bibitem{Cer05} V. Cerletti, O. Gywat, and D. Loss, cond-mat/0411235, version 2.
\bibitem{Ema05} C. Emary, B. Trauzettel, and C. W. J. Beenakker, cond-mat/0502550.

\end{thebibliography}
\end{document}